\documentclass[runningheads,citeauthoryear]{cl2emult}

\usepackage{makeidx}  
\usepackage{graphicx} 
\usepackage{subeqnar} 
\usepackage{multicol} 
\usepackage{cropmark} 
\usepackage{math}     
\makeindex            

\newcommand{\R}{{\bf     R}}
\newcommand{\N}{{\bf     N}}

\newcommand{\C}{{\bf     C}}
\newcommand{\E}{{\bf     E}}
\newcommand{\Prm}{{\bf     P}}

\newcommand{\e}{\varepsilon } 
\renewcommand{\epsilon}{\varepsilon } 
 
\renewcommand{\rho}{\varrho } 
 
\renewcommand{\phi}{\varphi }
\renewcommand{\a}{\alpha }
\renewcommand{\b}{\beta }
 
\renewcommand{\det}{{\rm det }}
\newcommand{\q}{{\rm q }}
\newcommand{\ran}{{\rm ran }}

\newcommand{\rs}{\right>}
\renewcommand{\lg}{\left|}

\begin{document}

\title{Optimal Summation and Integration by Deterministic, Randomized, and 
Quantum Algorithms}

\titlerunning{Deterministic, Randomized, and Quantum Algorithms}

\author{Stefan Heinrich\inst{1} \and Erich Novak\inst{2}}

\authorrunning{S. Heinrich, E. Novak}

\institute{Universit\"at Kaiserslautern, Fachbereich Informatik, 
Postfach 3049, D-67653 Kaiserslautern, \hbox{\tt heinrich@informatik.uni-kl.de}  
\and
Universit\"at Jena, Mathematisches Institut, 
Ernst-Abbe-Platz 4, D-07740 Jena, \hbox{\tt novak@mathematik.uni-jena.de}}

\maketitle              

\index{Heinrich@Stefan}
\index{Novak@Erich} 

\begin{abstract}
We survey old and new results about optimal algorithms for summation of 
finite sequences and for integration of functions from 
H\"older or Sobolev spaces.  First we discuss optimal deterministic and 
randomized algorithms. Then we add a new aspect, which has not been 
covered before on 
conferences about (quasi-) Monte Carlo methods: 
quantum computation. We give a short introduction
into this setting and
present recent results of the authors on optimal quantum algorithms
for summation and integration. We discuss comparisons between the three settings. 
The most interesting case for Monte Carlo and
quantum integration is that of moderate smoothness $k$ and large
dimension $d$
which, in fact, occurs in a number of important applied problems. In
that case the
deterministic exponent is negligible, so the
$n^{-1/2}$
Monte Carlo and the $n^{-1}$ quantum speedup essentially constitute the
entire convergence rate. We observe that 
\begin{itemize} 
\item 
there is an exponential speed-up of quantum algorithms over deterministic (classical)
algorithms, if $k/d$ tends to zero;
\item
there is a (roughly) quadratic speed-up of quantum algorithms over randomized
classical algorithms, if $k/d$ is small. 
\end{itemize} 
\end{abstract}

\section{Introduction} 

We study two numerical problems: Summation of finite sequences
and integration of functions. 
Besides the use of deterministic algorithms both are  prominent examples 
of the application of Monte Carlo methods and, quite
recently, also of quantum computation. When we study these problems from the point
of view of these three settings, we are usually interested in 
situations at the border of feasibility, that is, for the summation problem in a huge
number of summands $N$, and for the integration problem, in a large dimension $d$.
Clearly, the sum can be computed in $N-1$ additions exactly, but we 
want to understand if considerably fewer operations suffice
to provide an approximation of required precision. For integration it is clear that
we usually cannot compute the integral exactly, hence one has to settle for approximations.
For both problems the question arises of how good they can be solved in the given 
settings, and how the results compare. Such an approach incorporates two aspects:
Providing good algorithms  on one hand, 
and proving lower bounds -- that is, bounds such that no algorithm can be  
better -- on the other hand. Only when this is done in each of the settings, one can 
compare them on  theoretically founded grounds. Basic tools for such an analysis are 
provided by the theory of information-based complexity, see  Traub, Wasilkowski, 
Wo\'zniakowski (1988), Novak (1988), Heinrich (1993).

In this paper we want to 
give a survey of the state of the art in the above mentioned circle of questions.
Of course, comparisons between the deterministic and the randomized setting particularly
well fit the topic of this conference. On the other hand, we have also included the
new aspect of quantum computation. Although quantum computers exist so far
only in laboratories and their full computational power is still hypothetic,
many physicists and computer scientists believe in the future of quantum computing and
already now started to investigate, in which
problems a quantum computer can bring advantages over the classical one.
It is intriguing that quantum algorithms, 
although using a completely different model of computation, also possess and 
exploit features 
of randomization. So it is natural to study their relation to Monte Carlo methods.
In our investigations on quantum summation and integration, we found that certain
Monte Carlo techniques proved to be quite useful for the quantum setting. We discuss this 
below. Since this setting is certainly new to many people in the MCQMC community, we
chose to present it in a detailed way to give an insight 
not only into the special results but also into the 
general framework and model of computation.

Let us now formulate the problems under consideration. 
We are given a nonempty set $D$,  a set $F$ of real-valued functions on $D$ 
and an operator $S$ from $F$ to $\R$.  We seek to compute
(approximately) $S(f)$ for $f\in F$, where $f$ can only be accessed through its
values. In this paper we shall consider the summation operator (or mean operator) 
$S=S_N$
and the integration operator $S=I_d$, which are defined as follows:
For $f \in \R^N$ we define 
$$ 
S_N(f) = \frac{1}{N} \sum_{i=1}^N  f(i) 
$$ 
and for an integrable $f: [0,1]^d \to \R$ we define 
$$ 
I_d (f) = \int_{[0,1]^d}  f(x) \, dx \, . 
$$ 
We will study deterministic, randomized, and quantum algorithms 
for the computation of $S_N(f)$ and $I_d(f)$, up to some error $\e$. 

For the summation operator $S_N$ we assume that $f$ is from the set $F=L_p^N$ of all 
$f \in \R^N$ such that 
$$ 
\frac{1}{N} \sum_{i=1}^N  |f(i)|^p  \le 1 \,  ,
$$ 
$1 \le p < \infty$. (Let us mention that $L_p^N$ is the unit ball of the space $L_p(\mu)$, 
where $\mu$ is the equidistribution on $\{1,\ldots ,N\}$.) For $p=\infty$ we consider 
the set $L^N_\infty$ of 
all $f \in \R^N$ such that 
$$ 
|f(i)|  \le 1 \   \hbox{ for all } \  i \,  . 
$$ 

For the integration operator $I_d$ we assume 
that $f: [0,1]^d \to \R$ is from a H\"older class  
$F=F_d^{k,\a}$ or from a Sobolev class $F=W_{p,d}^k$. 
The H\"older classes are defined by 
\begin{eqnarray*}
\lefteqn{ 
F_d^{k,\a} =}\\
 &&\{ f \in C^k,\  \Vert f \Vert_\infty \le 1, 
  |D^i f (x) - D^i f (y) | \le |x - y|^\alpha , 
  \, x,y\in [0,1]^d,\ |i|=k  \} \,  , 
\end{eqnarray*}
where $k \in \N_0$, $0<\alpha \le 1$,  $C^k$ stands for the set of functions $f$ 
which are continuous together with all their partial derivatives $D^i f$ up to order $k$,
$\Vert \  \Vert_p$ ($1\le p \le\infty$) denotes the $L_p$-norm with respect to the 
Lebesgue measure on
$[0,1]^d$, $|x - y|$ is the Euclidean distance between $x$ and $y$,
 and $|i|$ means the sum of the components of the multiindex $i$. 
The Sobolev classes are defined by 
$$ 
W_{p,d}^k = \{ f  \,  : \,   \Vert D^i  f \Vert_p  \le 1,
 \ |i| \le k   \} \,  , 
$$ 
where $k \in \N_0$, $1 \le p \le \infty$, and $D^i$ is here the weak partial derivative.

For the integration problem in Sobolev spaces, we always assume the
embedding condition 
\begin{equation} \label{eq17} 
k \cdot p > d \,  , 
\end{equation}
which guarantees, by the Sobolev embedding theorem, that the elements of 
$W_{p,d}^k$ are continuous functions, and hence function values are well-defined.  

Let us briefly describe the organization of the paper. 
In Section 2 we survey known results about optimal 
deterministic algorithms for $S_N$ on $L_p^N$ and $I_d$ on $F_d^{k,\a}$ and 
$W_{p,d}^k$. Section 3 is concerned with randomized (or Monte Carlo) algorithms
for the same problems.  In Section 4 we give an introduction into the
model of quantum computation and survey recent results of 
Novak (2001) and Heinrich (2001a,b) 
on optimal algorithms for summation and integration on a quantum computer. 

\section{Deterministic Algorithms}

We  consider numerical algorithms of the form 
\begin{equation}   \label{eq8a} 
A_n(f) = \phi(f(x_1),\ldots,f(x_n))\, , 
\end{equation} 
where  $x_i\in D \,(i=1,\ldots,n),$ and $\phi:\R^n\to\R$ is an arbitrary mapping.
(In the terminology 
of information-based complexity, this is the class
of all nonadaptive, in general nonlinear algorithms using $n$  
function values.) A special subclass is formed by linear algorithms,
i.e.\  quadratures  
\begin{equation}   \label{eq8} 
A_n^{{\rm lin}}(f) = \sum_{i=1}^n  a_i \, f(x_i) 
\end{equation}
with $a_i\in\R$ and $x_i\in D$. 

The error of a method $A_n$ of the form (\ref{eq8a}) is defined as
$$ 
e(A_n,F) = \sup_{f \in F} |S(f) - A_n(f)| \, .
$$ 
The central quantity for our analysis is the $n$-th minimal error 
defined for  $n\in \N$ as\footnote{Formally one can also allow the case 
$n=0$ with constant algorithms $A_0$. For all the problems which we study 
one obtains $e_0^\det (F) =1$, hence all the problems are scaled in the same 
way.} 
$$ 
e_n^\det (F) = \inf_{A_n}  e(A_n,F) \, .
$$ 
The classes $L_p^N$, $F_d^{k,\a}$, and $W_{p,d}^k$ are unit balls in 
Banach spaces, so they are convex and symmetric. 
The operators $S_N$ and $I_d$ are linear. 
It is known that under these assumptions linear  methods 
(\ref{eq8}) are optimal (even among all adaptive, nonlinear methods). 
This was proved in Bakhvalov (1971), 
see also Novak (1996) and 
Traub, Wasilkowski, Wo\'zniakowski (1988). 
Therefore it is not difficult to find an optimal method 
for the summation operator $S_N$ on $L_p^N$
$$ 
A_n^* (f) = \frac{1}{N} \sum_{i=1}^n f(i) 
$$ 
and its error 
$$ 
e(A_n^*, L_p^N ) =  
\left( \frac{N-n}{N} \right)^{1-1/p} \,  , 
$$ 
where $n < N$. Of course we obtain 
$e_n^\det (L_p^N) = 0$ for $n \ge N$ and therefore always assume that $n < N$. 
The spaces $L_p^N$ are increasing with decreasing $p$ and for the extreme cases 
$p=\infty$ and $p=1$ we obtain 
$$ 
e_n^\det (L^N_\infty) = \frac{N-n}{N} \qquad \hbox{and} \qquad 
e_n^\det (L_1^N) = 1 \,  ,  \qquad n< N \, . 
$$ 
For later reference we summarize these (well known) results as follows. 
\begin{theorem}   \label{t1}    
Let $1 \le p \le \infty$ and $n < N$ . Then
$$ 
e_n^\det (L_p^N) = \left( \frac{N-n}{N} \right)^{1-1/p}. 
$$ 
\end{theorem} 
Now we discuss the integration problem. 
Again the results are classical, see Bakhvalov (1959) 
or references from information based complexity, 
such as  Novak (1988), 
Traub, Wasilkowski, Wo\'zniakowski (1988) or Heinrich (1993). 
We begin with the result for the H\"older classes.

\begin{theorem}   \label{t2} 
Let $k \in \N_0$ and $0<\alpha \le 1$. 
Then\footnote{
We write 
$a_n\asymp b_n$ iff there are $c_1,c_2>0$ and $n_0\in\N$ such that 
$c_1\, b_n\le a_n\le c_2\, b_n$ for all $n\ge n_0$. Similarly, we use
$a_n\prec b_n$ if there are $c>0$ and $n_0\in\N$ such that $a_n\le c\, b_n$
for all $n \ge n_0$. 
The respective constants $c$ or $c_i$ in this and the following statements 
may depend on the parameters
$k$, $\a$, $d$, and $p$, but do not depend on $n$ and $N$. 
}
$$ 
e_n^\det ( F_d^{k,\a} ) \asymp n^{-(k+\a)/d} \,  . 
$$ 
\end{theorem} 

\noindent For the Sobolev classes we have the following result. 

\begin{theorem}   \label{t3} 
Let  $k \in \N$ and $1 \le p \le \infty$, such that~(\ref{eq17}) holds. 
Then 
$$ 
e_n^\det ( W_{p,d}^k ) \asymp n^{-k/d} \,  . 
$$ 
\end{theorem} 

\section{Randomized Algorithms}

As numerical algorithms we now consider $A_n=(A_n^\omega)_{\omega\in \Omega}$,
where
\begin{equation}   \label{e21a}  
A_n^\omega (f) = \phi^\omega (f(x_1^\omega),\ldots, f(x_n^\omega)) \, ,  
\end{equation}
$(\Omega,\Sigma,\Prm)$ is a probability space, for each $\omega\in \Omega $,
$x_i^\omega\ $ is an element of $D$ and
$\phi^\omega $ is a mapping from $\R ^n$ to $\R$ with the property that
for each $f\in F$, the mapping 
$$
\omega\in\Omega\to A_n^\omega (f) = \phi^\omega (f(x_1^\omega),\ldots, f(x_n^\omega))
$$
is a random variable with values in $\R$. Again, randomized linear algorithms 
(randomized quadratures) are a special case, 
\begin{equation}   \label{e21} 
A_{n,\omega}^{{\rm lin}} (f) = \sum_{i=1}^n  a_i^\omega \, f(x_i^\omega) \, ,  
\end{equation} 
with random variables $a_i^\omega$ with values in $\R$  and $x_i^\omega$ 
with values in $D$ on $(\Omega,\Sigma,\Prm)$. 
The error of a method (\ref{e21a}) is 
\begin{equation}    \label{e21b} 
e(A_n^\omega ,F) = 
\sup_{f \in F} ( \E ( S(f) - A_n^\omega(f))^2)^{1/2} \, , 
\end{equation} 
where $\E$ is the expectation. 
The randomized $n$-th minimal error is defined as
$$ 
e_n^\ran (F) = \inf_{A_n}  e(A_n,F) \,  .
$$ 
Let us mention that in contrast to the deterministic setting no general result about 
the optimality
of linear methods among all methods is known for the randomized setting.
We start with the summation operator $S_N$ on $L_p^N$. 
Math\'e (1995) found the optimal randomized summation formula for 
$2 \le p \le \infty$. It has the following form: 
choose randomly an $n$-subset $\{i_1^\omega , \dots , i_n^\omega  \} \subset 
\{1, \dots N \}$, equidistributed on the family of all ${N\choose n}$
$n$-subsets and put 
$$ 
A_n^\omega (f) = c \,  \sum_{j=1}^n   f(i_j^\omega ) \, ,  
$$ 
where
$$ 
c= \left( n+ \sqrt{ \frac{n (N-n)}{N-1}} \right)^{-1} \, . 
$$ 
This $c$ satisfies 
$$ 
\frac{1}{n+\sqrt{n}} \le c \le \frac{1}{n} \, .
$$ 
Note also that here the optimal method is linear.  
We summarize the known results on 
$e_n^\ran (L_p^N)$ as follows (see Math\'e (1995)).

\begin{theorem}   \label{t4} 
Let $n < N$. Then
\begin{equation}   \label{eq28} 
e_n^\ran (L_p^N) = \frac{1}{1+ \sqrt{ \frac{(N-1) n}{N-n}}} \, , 
\qquad 2\le p \le \infty \, . 
\end{equation} 
For $1 \le p < 2$ only the order of the error is known, 
\begin{equation} \label{eq28a}
e_n^\ran (L_p^N) \asymp  n^{-1+1/p}  \, , 
\end{equation} 
for $N > \beta n$, where $\beta > 1$. 
\end{theorem} 
In particular it follows from~(\ref{eq28}) that 
$$ 
e_n^\ran (L_p^N) \asymp  n^{-1/2}    \, , 
$$ 
for $2 \le p \le \infty$ and $ N > \beta n$, where $\beta >1$.

We want to mention a couple of 
interesting facts  around relation (\ref{eq28a}).
One might ask if the classical Monte Carlo method (sampling independently, 
uniformly on $\{1,\ldots,N\}$) yields this optimal rate for $1<p<2$.
It does not. The reason is that the variances of functions
in $L_p^N$, which we need for the
error as defined in (\ref{e21b}), will not be bounded uniformly in $N$.
Even more is true: no linear methods (\ref{e21}) can reach this rate. 
Math\'e (1992) proved that for 
$1 \le p < 2$ and $ N > \beta n$ the error of optimal linear methods is of the order 
$$
  \min (N^{1/p-1/2}n^{-1/2},1)\, .
$$ 
However, it is easily checked that
a slight (nonlinear) modification of the classical Monte Carlo method 
does give the optimal rate: replace the function $f\in L_p^N$ by
$\tilde{f}$ defined by $\tilde{f}(i)=f(i)$ if $|f(i)|\le n^{1/p}$
and $\tilde{f}(i)=0$ otherwise. Then apply standard Monte Carlo to $\tilde{f}$.
It is also interesting that these results are sensitive with respect to
the error criterion (\ref{e21b}). If we replace the $L_2$ norm
$$
(\E ( S_N(f) - A_n^\omega(f))^2)^{1/2}
$$
by the $L_p$ norm 
$$
(\E ( S_N(f) - A_n^\omega(f))^p)^{1/p}
$$
(or any $L_q$ norm with $1\le q\le p < 2$), classical Monte Carlo (a linear method)
does provide the optimal rate -- the same rate as (\ref{eq28a}).  
See Heinrich (1993), where this is shown for $q=1$, but the proof gives also 
$q=p$ (from which $q<p$ follows trivially).
Now we discuss the integration problem. 
For the H\"older classes we have the following result from Bakhvalov (1959).

\begin{theorem}   \label{t7} 
Let $k \in \N_0$ and $0<\alpha \le 1$. 
Then 
$$ 
e_n^\ran ( F_d^{k,\a} ) \asymp n^{-(k+\a)/d -1/2} \,  . 
$$ 
\end{theorem} 
The results for the Sobolev classes are again due to Bakhvalov (1962), 
see also Novak (1988) and Heinrich (1993). 
\begin{theorem}   \label{t8} 
Let $k \in \N_0$, $1 \le p \le \infty$, and 
assume that the embedding condition~(\ref{eq17}) holds.  Then 
$$ 
e_n^\ran ( W_{p,d}^k ) \asymp n^{-k/d - 1/2 } \quad  \hbox{for}\ 2\le p \le \infty \,  , 
$$
and
$$
e_n^\ran ( W_{p,d}^k ) \asymp n^{-k/d - 1+1/p } \quad \hbox{for} \ 1\le p < 2 \,  .
$$
\end{theorem}

\section{Quantum Algorithms}

By now it is well-known that quantum computers (if one succeeded in building them)
could yield considerable speed-ups for certain important discrete 
problems. Shor's (1994, 1998) quantum algorithm  for factorization of 
integers is polynomial in the number of 
bits -- while no such algorithm for a classical
computer is known, and, moreover, nowadays secret codes firmly rely on the hope 
that no such algorithms exist. 
Grover (1996, 1998) presented a quantum search algorithm which
finds a specified  element out of $N$ items in $\Theta(\sqrt{N})$ operations, whereas
classically $\Theta (N)$ are necessary.
These two discoveries largely encouraged quantum computer research and triggered a 
stream of 
further investigations into the potential powers of quantum computers. Up to now,
however, this research dealt almost exclusively with discrete problems.
Here we want to know whether quantum computers are useful for 
problems like summation of reals or integrals. 

First we introduce the model of computation. For basic notions, background,
and further material on quantum computing we refer to the surveys and monographs: 
Ekert, Hayden, Inamori (2000), Shor (2000), Pittenger (1999), Gruska (1999) and
Nielsen, Chuang (2000).
Let $H_1$ be a 2-dimensional Hilbert space over $\C$ and let 
$e_0$ and $e_1$ be two orthonormal vectors in $H_1$. 
The space $H_1$ represents a quantum bit, 
in the Dirac notation we have 
$$
e_0 = \lg 0 \rs \quad \hbox{and} \quad e_1 = \lg 1 \rs \,  .
$$
For $m \in \N$ quantum bits we use the $2^m$-dimensional tensor product space
$$
H_m= H_1 \otimes \dots \otimes H_1
$$
with $m$ factors. An orthonormal basis is given by the $2^m$ vectors 
$$
b_\ell  = e_{i_1} \otimes \dots \otimes e_{i_m} \,  , 
$$
where $i_j \in \{ 0, \, 1 \}$ and 
$$  
\ell = \sum_{j =1}^m i_j \, 2^{m-j} , \qquad \ell =0, \dots , 2^m -1 \, . 
$$ 
There are $2^m$ different $b_\ell $ and this corresponds 
to the $2^m$ different possibilities 
of an information that is given by $m$ classical bits. 
The Dirac notation for $b_\ell $ is just $\lg \ell  \rs$, instead of 
$e_{i_1} \otimes e_{i_2}$ one finds $\lg i_1, i_2 \rs$ or also 
$\lg i_1 \rs \lg i_2 \rs$. 
The formally different objects $(i_1, \dots , i_m)$ and $\ell$ or $b_\ell$ are often 
identified and called classical state or basis state. 
Let $\mathcal{U}(H_m)$ denote the set 
of unitary operators on $H_m$.

The decomposition of  $x \in H_m$ with respect to the basis $(b_\ell)$ is given by
$$ 
x = \sum_{i_j \in \{ 0,1\} } \a_{(i_1, \dots , i_m)} e_{i_1} 
\otimes \dots \otimes e_{i_m} =
\sum_{\ell=0}^{2^m-1} \alpha_\ell \, b_\ell \,  .
$$ 
We are only interested in normed vectors, $\Vert x \Vert =1$. 
All such vectors are called (pure) ``quantum states''. 
For each quantum state there is a probability distribution on the classical states: 
the probability of $\ell$ is $|\alpha_\ell|^2$. This is a typical feature of
quantum algorithms -- we cannot ``read'' (measure) the coordinates $ \alpha_\ell$
of a state $x$ as above. When we measure $x$, the result is probabilistic:
we obtain a classical state $\lg \ell\rs$ with probability $ |\alpha_\ell|^2$.

Now let, as in the previous sections,  
$D$ be a nonempty set,  $F$ a set of real-valued functions on $D$ 
and let  $S$ be an operator from $F$ to $\R$.  We describe in the sequel what we
mean by a quantum algorithm on $m$ qubits for the (approximate) computation of $S$.
For this purpose we first need to introduce the notion of a quantum query.
We follow Heinrich (2001a). Like
the classical algorithms get information about the input $f$ through 
function values, a quantum algorithm can access $f$ through a quantum query
$$
Q=(m',m'', Z, \tau,\b) \, ,
$$
where $m',m''\in\N$, $m'+m''\le m$ with $m$ as above (the number of qubits),
$Z$ is a nonempty subset of $\{0,\ldots,2^{m'}-1\} ,$ and 
$$
\tau:Z\to D
$$
and 
$$
\b:\R\to\{0,\ldots,2^{m''}-1\}
$$
are any mappings.
The meaning of these components is the following: $\tau$ 
is the rule by which we associate to a bit-string $i\in Z$ the node point 
$\tau (i)\in D$ at which
we want to evaluate $f$ (we include the case that $\tau$ is not defined on
all of $\{0,\ldots,2^{m'}-1\} ,$ which is convenient e.g.\ if the number
of elements of $D$ is finite, but not a power of 2). When we have $f(\tau(i))$, 
which is an element of $\R$,
we still need to convert it into a binary string -- which is the r\^{o}le
of the mapping $\b$. 

With a quantum query $Q$ and an input $f\in F$ we associate
a unitary operator $Q_f\in \mathcal{U}(H_m)$ which is defined on the basis state
$$\lg i\rs\lg x\rs\lg y\rs\in
H_{m}=H_{m'}\otimes H_{m''}\otimes H_{m-m'-m''}
$$
as
$$ 
Q_f\lg i\rs\lg x\rs\lg y\rs=
\left\{\begin{array}{ll}
\lg i\rs\lg x\oplus\beta(f(\tau(i)))\rs\lg y\rs &\quad \mbox {if} \quad i\in Z\\
\lg i\rs\lg x\rs\lg y\rs & \quad\mbox{otherwise}\, . 
 \end{array}
\right. 
$$
Here $\oplus$ denotes addition modulo $2^{m''}$. It is easily 
seen that $Q_f$ is a bijection on the set of basis states, 
and hence can be extended uniquely to a unitary operator on $H_m$.

A quantum algorithm with $n$ quantum queries is a tuple
$$
A_n=(m,w,Q,(U_i)_{i=0}^n,\varphi) \, ,
$$
where $m$ is the number of qubits (i.e.\ the algorithm acts on $H_m$), 
$w\in H_m$ is a basis state (the starting state of the algorithm),
$Q$ is a quantum query as defined above (supplying the information about $f$),
$U_i\in \mathcal{U}(H_m)$ are any fixed unitary operators (the quantum
computations), and
$$
\varphi:\{0,\ldots,2^m-1\}\to\R
$$
is an arbitrary mapping.
The mapping $\phi$ produces the real number which is the 
output of the quantum computation. The algorithm $A_n$ acts on input $f$ as follows:
It starts with the state $w$, to which $U_0$ is applied, which gives
$ U_0 w.$
The mapping $U_0$ unites all quantum operations before the first query call. 
Then $Q_f$ is applied, leading to
$Q_f U_0 w.$
 Next $U_1$ is 
applied (standing for the quantum operations between the first and the
second query call), yielding $U_1 Q_f U_0 w.$
Then $Q_f$ is called again, etc. In the end the algorithm produces
the state
$$
z=U_n Q_f U_{n-1}\ldots U_1 Q_f U_0 w \, .
$$
Let
$$      
z=\sum_{\ell=0}^{2^m-1} \alpha_{\ell,f} \lg \ell\rs \,  .
$$
As we mentioned before, we cannot access the components $\alpha_{\ell,f}$ of
$z$ directly. The state $z$ is measured, giving a 
random variable $\xi_f(\omega)$ with values in
$\{0,\ldots ,2^m-1\}$, which takes the value $\ell$ with probability
$|\alpha_{\ell,f}|^2$. 
Finally the mapping $\phi$ is applied, which
stands for the computations performed on the result of the measurement
(on a classical computer). So the output of the algorithm is 
$$
A_n(f,\omega)=\phi(\xi_f(\omega)) \, .
$$

Note two important things: Firstly, the algorithm gets information 
about $f$ only through $Q_f$, while the state $w$, the unitary operators $U_i$ and 
the mapping $\phi$ are fixed from 
the beginning and do not depend on $f$. Secondly, each $U_i$ stands, in fact, for 
a sequence of elementary quantum operations, called gates (like the basic
logical operations in classical computation).  
The gates can be chosen in such a way that each unitary operator can
be represented as a finite compositions of gates.  
For more details we refer to Ekert, Hayden, Inamori (2000). Here we are
concerned with the query complexity, meaning that 
we want to study the minimal error which can be reached by algorithms using 
at most $n$ queries (this is essentially parallel to the previous sections).
So the number of gates needed to represent (implement) the $U_i$ will not
be discussed.

What we described above is a quantum algorithm with a single 
measurement. One can put together several such 
algorithms to obtain an algorithm with multiple measurements: 
the output of the first part of the algorithm is used (by a classical computer)
to compute the input of the second part, and so on.
A formal description of this is given in Heinrich (2001a), where
it is also shown that from the point of view of query complexity it makes  
(up to a factor of 2) no difference if algorithms with one or with several measurements
are considered. 

The error of $A_n$ at input $f$ is defined probabilistically as
$$
e(A_n,f)=\inf\left\{\varepsilon:\,\Prm\{|S(f)-A_n(f,\omega)|\le\varepsilon\}\ge 
3/4 \right\} \, ,
$$
and 
$$
e(A_n,F)=\sup_{f\in F} e(A_n,f) \, . 
$$
Note that we require an error probability not greater than 1/4. By 
repeating the algorithm $k$ times and computing the median of the results, 
the error probability can be reduced to $2^{-ck}$ for some $c>0$ not depending on $k$.
Finally we define
$$
e_n^\q (F) = \inf_{A_n}  e(A_n,F) \, .
$$

Now we study  the numbers $e_n^\q(L_p^N)$, 
$e_n^\q(F_d^{k,\alpha})$ and 
$e_n^\q(W^k_{p,d})$.
Again we start with the summation operator $S_N$ on $L_p^N$.
The following result for $1<p<\infty$ is from Heinrich (2001a), 
in the case $p=\infty$ 
the upper bound is due to Brassard, H\o yer, Mosca, Tapp (2000), 
the lower bound due to Nayak, Wu (1998).

\begin{theorem}   \label{t9} 
Let $n<N$. Then
$$
e_n^\q (L_p^N)\asymp n^{-1} \quad \hbox{for} \ 2<p \le \infty \,  ,
$$
$$
n^{-1}\prec e_n^\q (L_2^N) \prec n^{-1} \log^{3/2} n \,  \log \log n \, ,
$$
and  
$$
e_n^\q (L_p^N) \asymp n^{-2+2/p} \quad \hbox{for} \ 1<p <2 \, ,
$$
where the lower bound  for $e_n^\q (L_p^N)$ in the last relation holds under 
the restriction $n^2\le N$.
\end{theorem} 
Next we discuss the integration problem.
For the H\"older classes we have the following result from 
Novak (2001). 

\begin{theorem}   \label{t12} 
Let $k \in \N_0$ and $0<\alpha \le 1$. 
Then 
$$ 
e_n^\q ( F_d^{k,\a} ) \asymp n^{-(k+\a)/d -1} \, . 
$$ 
\end{theorem} 

\noindent 
For the Sobolev classes Heinrich (2001b) proved the following result. 

\begin{theorem}   \label{t13} 
Let $k \in \N_0$ and $1 \le p \le \infty$.  
If the embedding condition~(\ref{eq17}) holds, then
$$ 
e_n^\q ( W_{p,d}^k) \asymp n^{-k/d -1} 
\quad \hbox{if} \quad 2<p \le \infty \,  ,
$$ 
$$ 
n^{-k/d -1}\prec e_n^\q ( W_{2,d}^k) \prec
n^{-k/d-1}\,\log^{3/2}n\log \log n \, ,
$$ 
and
$$ 
e_n^\q ( W_{p,d}^k)  
\prec  n^{-k/d-3/2+1/p}\,\log^{3/2}n\log \log n
$$  
if $1 \le p < 2$.
\end{theorem} 

\noindent 
We do not know the optimal order in the case $1 \le p <2$, 
for $p=2$ there is a logarithmic gap between the upper 
and lower bounds. 
Another gap is in the summation problem with 
$1 < p< 2$ and relatively small $N$. Summation in the case $p=1$
is open, as well. These questions are the subject of ongoing 
research.

Let us finally present the results in a table and discuss some comparisons.
In the following  we omit log-factors and exhibit the asymptotic order 
of the error for $0<n<N/2$.

It can be read from the table that  in the case  
$2\le p\le\infty$
Monte Carlo gives a speedup of $n^{-1/2}$ over deterministic algorithms both in
summation and integration. In these
cases quantum algorithms give a speedup of $n^{-1}$ (over deterministic algorithms),
that is, the Monte Carlo gain
is squared. A similar pattern can be found in the case 
$1< p<2$ for summation of sequences.
The speedup of Monte Carlo is here $n^{-1+1/p}$  
(hence the advantage over deterministic algorithms is reduced as $p$ decreases).
Nevertheless, in the quantum setting this speedup is squared again.
In the case
$1\le p<2$ for integration in Sobolev spaces, the quantum gain is still better:
Even in the case $p=1$, where Monte Carlo gives no advantage at all over 
deterministic algorithms, the quantum speedup is  $n^{-1/2}$. 

\renewcommand{\thefootnote}{\fnsymbol{footnote}}
\[ \renewcommand{\arraystretch}{1.5}
\begin{array}{l|l|l|l}
& \ \mbox{deterministic}\  & \quad \mbox{randomized}\, & \quad \mbox{quantum}\,\\ \hline
L_p^N,\,2\le p\le \infty &\quad 1& \quad n^{-1/2}        & \quad n^{-1}\\
L_p^N,\,1< p<2 & \quad 1&\quad  n^{-1+1/p} & \quad  n^{-2+2/p}  \,\quad 
\footnotemark[1]\\
F_d^{k,\alpha} &\quad n^{-(k+\a)/d}  &\quad  n^{-(k+\a)/d-1/2 \  } 
& \quad  n^{-(k+\a)/d-1}\\
W_{p,d}^k,\,2\le p\le \infty &\quad n^{-k/d } & \quad n^{-k/d -1/2 }  
& \quad n^{-k/d -1}\\
W_{p,d}^k,\,1\le p<2 &\quad n^{-k/d } & \quad n^{-k/d -1 + 1/p }
& \quad n^{-k/d -3/2 + 1/p }\quad\footnotemark[2]\\
\end{array}
\]
\footnotetext[1]{lower bound only for $N\ge n^2$}
\footnotetext[2]{only upper bound}
\renewcommand{\thefootnote}{\arabic{footnote}}

The most interesting case for Monte Carlo and
quantum integration is that of moderate smoothness
$k$ and large dimension%
\footnote{For the Sobolev
spaces, as a consequence of the embedding
condition (\ref{eq17}), also $p$ has to be appropriately large.}
$d$ which, in fact, occurs in a number of important applied problems. In
that case the
deterministic exponent, $(k+\alpha)/d$ or $k/d$, is negligible, so the
$n^{-1/2}$
Monte Carlo and the $n^{-1}$ quantum speedup essentially constitute the
entire convergence rate. 
Hence we observe a situation similar to that of
Grover's
search algorithm: a quadratic speedup for quantum computation as
compared to
classical randomized algorithms.
If we compare quantum algorithms with deterministic classical algorithms, 
then the speed-up is even much larger -- it is exponential in the dimension $d$.

\end{document}